\documentclass[twocolumn,showpacs,preprintnumbers,amsmath,amssymb,superscriptaddress]{revtex4-2}

\usepackage{graphicx}% Include figure files
\usepackage{dcolumn}% Align table columns on decimal point
\usepackage{amsmath}
\usepackage{amssymb}
\usepackage{bm}% bold math
\usepackage{soul} % strikethough, highlight
\usepackage{psfrag}
\usepackage{float}
\usepackage{subeqnarray}
\usepackage{xcolor}
\usepackage{multirow}
\usepackage[normalem]{ulem}

\usepackage{url}
\usepackage[colorlinks=true,linkcolor=blue,citecolor=blue,urlcolor=blue,breaklinks]{hyperref}
\begin{document}

	\title{Three-component superconductivity: the effect of second-order Josephson couplings}

	\author{Shen-Yi Peng}
	\affiliation{Department of Physics, and Shanghai Key Laboratory of High Temperature Superconductors, Shanghai University; Shanghai 200444, China.}
	\author{Ling-Feng Zhang}%
	\email{lingfeng\_zhang@shu.edu.edu}
	\affiliation{Department of Physics, and Shanghai Key Laboratory of High Temperature Superconductors, Shanghai University; Shanghai 200444, China.}
	\author{Xiao Hu}
    \affiliation{Department of Physics, and Shanghai Key Laboratory of High Temperature Superconductors, Shanghai University; Shanghai 200444, China.}
	\affiliation{Institute for Quantum Science and Technology, Shanghai University; Shanghai 200444, China.}

	\date{\today}

	\begin{abstract}
		Recently, a three-component Ginzburg-Landau (GL) model compatible with the 3Q pair-density-wave state has been proposed to explain the fractional quantum magnetic resistance oscillations of period $\phi_0/3 = hc/6e$ observed in vanadium-based kagome superconductors. The physics of this model is governed by second-order Josephson-type couplings, which break both time-reversal symmetry and discrete $\pi$-phase flip symmetry.  In this work, we theoretically derive the complete set of ground-state solutions and construct a comprehensive phase diagram in the GL parameter space, characterized by analytically determined phase boundaries.  We identify five distinct ground states: an 8-fold degenerate frustrated state and four 4-fold degenerate non-frustrated phase-locked states. Four of these states spontaneously break time-reversal symmetry. Numerical analysis of the collective modes reveals the emergence of a Higgs-Leggett mode unique to the frustrated region, accompanied by mode softening near the phase boundaries. Our findings provide a comprehensive theoretical framework for understanding the multifaceted physics of multicomponent superconductivity.
	\end{abstract}
	
	
	\maketitle
	%\tableofcontents

% #################################################
% #########  SECTION I #######################
% #################################################
\section{Introduction}

Multiband superconductivity and the induced unconventional quantum states have long been a frontier in condensed matter physics\cite{Agterberg_AnnualReviews_2020, PhysRevResearch.7.023129, Tanaka_2015,Lin_2014,Zehetmayer_2013,jhy9-mckg,7hhv-9fq7,k2t5-wcq1,PhysRevB.84.214505}.  Initially, the surge of interest in three-component superconducting systems was primarily driven by the discovery of iron-based superconductors\cite{Kamihara_2006}, where the multiband structure not only explains the celebrated $s_{\pm}$ pairing symmetry but also predicts exotic $s+is$ states\cite{PhysRevLett.101.057003,Chubukov_2012}.  
%doi:10.7566/JPSCP.1.012121,PhysRevB.83.054515,
Within a three-band framework, first-order Josephson coupling can induce frustration among the order parameter phases, leading to spontaneous time-reversal symmetry breaking (TRSB) \cite{Tanaka_2010,Stanev_2010,Hu_2012}.  This frustration-induced state naturally hosts a rich variety of collective excitations and topological defects, including Higgs-Leggett collective modes \cite{PhysRevLett.108.177005,PhysRevB.84.134518}, fractional magnetic flux vortices \cite{PhysRevB.92.214516}, domain walls \cite{PhysRevLett.112.017003} and topological solitons where closed domain walls bind fractional vortices\cite{Lin_2012,PhysRevLett.107.197001,PhysRevB.87.014507,PhysRevB.107.094503,PhysRevB.104.014518}.  Experimental evidence for vortex fractionalization was reported on the potassium-terminated surface of the multiband superconductor $\text{KFe}_2\text{As}_2$ \cite{Zheng_science_2026}.

The recent observation of $\phi_0/3=hc/6e$ fractional flux quantization in kagome superconductors $\mathrm{CsV_3Sb_5}$, necessitates a fundamental re-examination of these existing frameworks \cite{PhysRevX.14.021025,Zhang_2024}.   As a versatile platform for exploring the interplay among geometric frustration, topology, and strong correlations, kagome superconductors manifest a rich spectrum of exotic phenomena \cite{PhysRevMaterials.3.094407,jiang_kagome_2022,yin_topological_2022,neupert_charge_2022}. These include van Hove singularities \cite{hu_rich_2022}, flat-band physics \cite{PhysRevB.111.155123}, and various intertwined orders such as charge-density waves (CDW)\cite{nie_charge-density-wave-driven_2022,Han_2023}. 

Distinct from iron-based systems, the unique geometric symmetries of kagome superconductors like $\mathrm{CsV_3Sb_5}$ support a $3Q$ pair-density-wave (PDW) order \cite{Chen_2021,Zhao_2021}.  In this framework, the superconducting order parameter comprises three components associated with the triple-$Q$ directions, each characterized by identical physical properties.  Notably, momentum conservation strictly suppresses conventional first-order Josephson coupling within the PDW state, establishing second-order Josephson coupling as the leading-order interaction \cite{PhysRevLett.129.167001}. While a repulsive second-order term similarly induces a TRSB state, its physical consequences are far more intricate than those of its first-order counterpart, ultimately yielding a ground-state with 8-fold degeneracy \cite{ZHANG20241354512}. 

To provide a comprehensive understanding of the physical impact of second-order Josephson coupling, this paper presents a systematic investigation of the model's expansive parameter space. We focus on the ground-state properties, their stable parameter ranges and collective excitations.  By obtaining analytical solutions for the ground states, we construct a detailed phase diagram and provide exact expressions for the phase boundaries and stability conditions. Within a linear-response framework, we calculate the collective excitation spectra and their characteristic coherence lengths.  We identify soft modes at the phase boundaries, where the closing of the excitation gap directly confirms our analytical phase-transition predictions. Our findings are applicable not only to ideally symmetric Kagome superconductors but also to broader classes of multiband systems exhibiting stress-induced anisotropy or unequal components.

The paper is organized as follows: Sec.~\ref{sec:model} introduces the three-component Ginzburg-Landau (GL) model with second-order Josephson coupling; Sec.~\ref{subsec1} derives the analytical ground-state solutions; Sec.~\ref{subsec2} constructs the phase diagram in GL parameter space and determines the phase boundaries; Sec.~\ref{subsec3} calculates the collective excitation spectra and coherence lengths under linear response; Finally, Sec.~\ref{sec:conclusion} summarizes our findings.

% #################################################
% #########  SECTION II #######################
% #################################################
\section{The Ginzburg-Landau Model}
\label{sec:model}

In the absence of an external magnetic field, the three-component GL free energy density functional is expressed as\cite{Zhang_2024}:
\begin{equation}
    \begin{split}
        G &= \sum_{j=1,2,3} \left[ a_j |\psi_j|^2 + \frac{b_j}{2} |\psi_j|^4 + \frac{\hbar^2}{2m_j} \left| \nabla \psi_j \right|^2 \right] \\
        &- \sum_{\substack{j,k=1,2,3 \\ j<k}} \frac{\eta_{jk}}{2} \left( \psi_j^{*2} \psi_k^2 + \text{c. c.} \right),
    \end{split}
    \label{eq:Gall}	
\end{equation}
where the three order parameters are defined as $\psi_j = \sqrt{n_j} e^{i\theta_j}$ for $j=1, 2, 3$. The central feature of this model is the second-order Josephson-coupling term, characterized by the parameters $\eta_{jk}$.  This interaction term can be explicitly written as $-\eta_{jk} n_j n_k \cos(2\theta_{kj})$, where $\theta_{kj} = \theta_k - \theta_j$.  

The GL free energy $G$ is invariant under the continuous $U(1)$ gauge transformation. Beyond this, the second-order Josephson coupling introduces two critical discrete symmetries. First, the functional is invariant under time-reversal symmetry ($\mathcal{T}$), defined by $\psi_j \to \psi_j^*$ (or $\theta_{kj} \to -\theta_{kj}$), the spontaneous breaking of which leads to a TRSB state. Second, the $2\theta$ dependence of the Josephson terms grants the system a discrete $\pi$-phase flip symmetry, defined by $\psi_j \to -\psi_j$ (or $\theta_{jk} \to \theta_{jk} + \pi$). The interplay between these discrete symmetries facilitates the emergence of non-trivial frustrated configurations within the ground-state manifold.

The interplay between these symmetries dictates the degeneracy and structure of the ground-state. For a repulsive interaction ($\eta_{jk} < 0$), the energy is minimized at a relative phase difference of $\theta_{jk} = \pm \pi/2$, while an attractive interaction ($\eta_{jk} > 0$) favors a collinear alignment of $\theta_{jk} = 0 \pmod{\pi}$. The competition among these couplings leads to a rich variety of ground-state configurations driven by phase frustration, most notably the 8-fold degenerate regime where both symmetries are spontaneously broken.

For simplicity, we consider the case where all components share an identical transition temperature ($T_{c1} = T_{c2} = T_{c3} = T_c$). Under this condition, all three order parameter components emerge simultaneously at the transition point. This choice allows us to focus on the three-component regime while avoiding sequential symmetry breaking, which typically leads to intermediate single-component \cite{PhysRevLett.78.2208, PhysRevLett.81.2783, PhysRevLett.97.137002, PhysRevLett.103.267002}  or two-component \cite{PhysRevB.91.161102, PhysRevB.86.060514, PhysRevB.93.014518, PhysRevB.94.064519, PhysRevB.101.064501, PhysRevLett.106.047005} superconducting phases. Hereafter, the parameters are given by:
\begin{equation}\label{eq:a_coeffs_3line}
        a_j = a_{j0} (1 - T/T_c), \quad j=1,2,3
\end{equation}
where $a_{j0} < 0$. As a result, all components satisfy $|\psi_j| > 0$ for $T < T_c$, while the system remains in the normal state ($|\psi_j| = 0$) for $T > T_c$.

% #################################################
% #########  SECTION III #######################
% #################################################
\section{Result}
\label{sec:result}

% #################################################
% ######## subsection I  ##########################
\subsection{Analytic Ground-state Solutions}
\label{subsec1}

\begin{figure}
    \includegraphics[width=0.8\columnwidth]{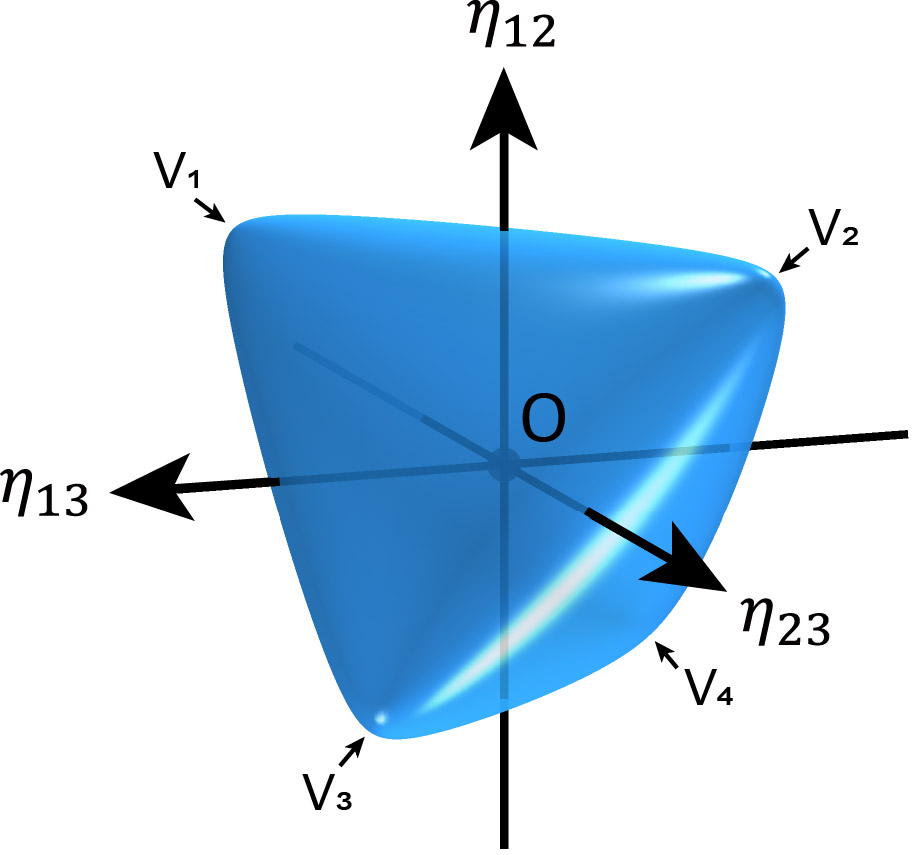}
    \caption{\textbf{Schematic stability domain of the three-component superconducting states.} The closed, curved blue tetrahedral region in the $(\eta_{12}, \eta_{13}, \eta_{23})$ parameter space defines the stability domain: the three-component superconducting state is stable within this region, where the matrix $\mathbf{T}$ is positive definite ($\mathbf{T} > 0$). The coordinates of its four vertices, $V_1$--$V_4$, are given in Eq.~\eqref{eq:v1v2v3v4}.}
    \label{fig: tetrahedron}
\end{figure}

\begin{figure}
    \includegraphics[width=0.8\columnwidth]{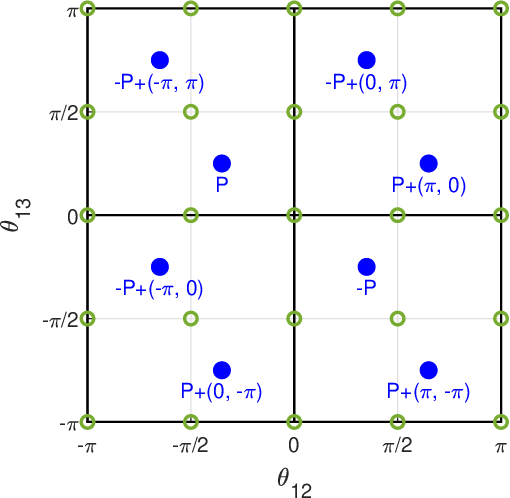}
    \caption{\textbf{Distribution of phase difference solutions in the $(\theta_{12}, \theta_{13})$ plane} The relative phase solutions $(\theta_{12}=\theta_1-\theta_2, \theta_{13}=\theta_1-\theta_3)$ for the Ginzburg-Landau equations are determined by Eq.~\eqref{eq:dtheta}.  For a given set of parameters, there exist a maximum of 24 distinct solutions, categorized into two groups based on their symmetry properties.  Sixteen of these solutions (open green circles) are pinned at high-symmetry points (e.g., multiples of $\pi/2$), representing the non-frustrated, phase-locked configurations.  The remaining eight solutions (solid blue dots) are located at low-symmetry points and constitute the 8-fold degenerate frustrated state.  Given a representative solution $P$ defined by the relative phases in Eq.~\eqref{eq:phase_solutions_final}, the other seven counterparts are generated via symmetry operations: $-P$ is obtained through inversion symmetry, while the others are generated by discrete $\pi$-translations in $\theta_{12}$, $\theta_{13}$, or both.
    }
    \label{solutions1}
\end{figure}

\begin{figure}
    \includegraphics[width=0.7\columnwidth]{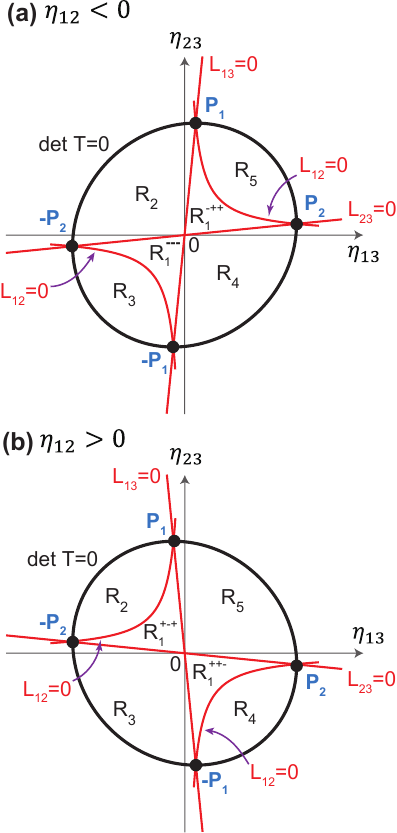}
    \caption{\textbf{Topological partition of the frustrated state in the $(\eta_{12}, \eta_{13}, \eta_{23})$ parameter space.}  Panels (a) and (b) illustrate the $(\eta_{13}, \eta_{23})$ cross-sections for fixed $\eta_{12} < 0$ and $\eta_{12} > 0$, respectively.  The solid black ellipse denotes the stability threshold $\det \mathbf{T} = 0$, defining the stable domain for the three-component superconducting state. Within this stable domain, the 3D parameter space is partitioned by the conditions $L_{jk} = 0$ (solid red lines) into eight distinct sectors: the four frustrated sectors ($R_1^{---}, R_1^{-++}, R_1^{+-+}, R_1^{++-}$) and four non-frustrated sectors ($R_2$--$R_5$). For the frustrated sectors, the superscripts denote the signs of ($L_{12}$, $L_{13}$, $L_{23}$).  These sectors satisfy the condition $L_{12}L_{13}L_{23}<0$.  High-symmetry intersection points $P_1$, $P_2$, and their inversion-symmetric counterparts $-P_1$ and $-P_2$ are marked, with coordinates given by Eq.~\eqref{eq:P1P2}.
    }
    \label{fig:LLL}
\end{figure}

\begin{figure}
    \includegraphics[width=\columnwidth]{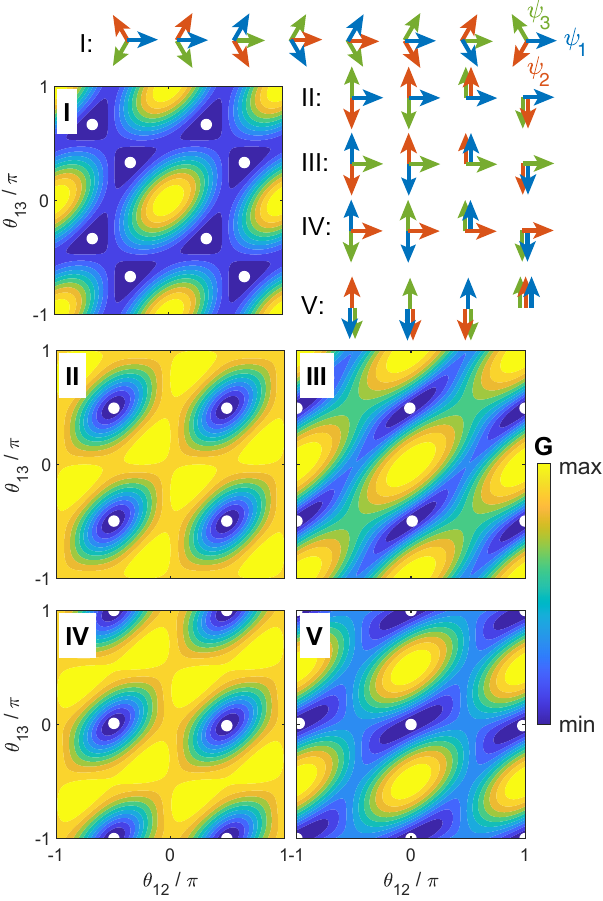}
    \caption{\textbf{Classification of ground-state configurations.}  The ground-state solutions are categorized into five distinct regimes (I--V) determined by the Ginzburg-Landau parameters.  The contour plots display the corresponding minimal free-energy landscapes $G(\theta_{12}, \theta_{13})$,  where the white dots mark the global energy minima.  (I) An 8-fold degenerate ground-state (GS) with minima located at general points in the phase plane.   (II--V) 4-fold degenerate ground-state configurations where the phase differences are pinned at high-symmetry points. The vector diagrams (upper right) illustrate the relative orientations of the three-component order parameters $(\psi_1, \psi_2, \psi_3)$ for each regime. Note that for regime I, the vector diagrams are representative examples, as the equilibrium phase differences vary continuously with model parameters.}
    \label{solutions2}
\end{figure}

The ground-state is determined by the stationary points of the GL free energy $G$, where kinetic energy contributions are neglected. By taking the functional variation of $G$ in Eq.~\eqref{eq:Gall} with respect to the order parameters $\psi_j^*$ (i.e., $\delta G / \delta \psi_j^* = 0$ for $j=1, 2, 3$), we obtain the following set of coupled GL equations:
\begin{equation}
    \begin{cases}
        a_1 \psi_1 + b_1 |\psi_1|^2 \psi_1 - \eta_{12} \psi_2^2 \psi_1^* - \eta_{13} \psi_3^2 \psi_1^* = 0 \\
        a_2 \psi_2 + b_2 |\psi_2|^2 \psi_2 - \eta_{12} \psi_1^2 \psi_2^* - \eta_{23} \psi_3^2 \psi_2^* = 0 \\
        a_3 \psi_3 + b_3 |\psi_3|^2 \psi_3 - \eta_{13} \psi_1^2 \psi_3^* - \eta_{23} \psi_2^2 \psi_3^* = 0.
    \end{cases}
    \label{eq:ground1}
\end{equation}

To find the solutions representing possible ground-state configurations, we decompose the order parameters into their respective superfluid densities and phases:
\begin{equation}\label{eq:ground2}
    \begin{cases}
        a_1 + b_1 n_1 - \eta_{12} n_2 e^{i2\theta_{21}} - \eta_{13} n_3 e^{i2\theta_{31}} = 0 \\
        a_2 + b_2 n_2 - \eta_{12} n_1 e^{i2\theta_{12}} - \eta_{23} n_3 e^{i2\theta_{32}} = 0 \\
        a_3 + b_3 n_3 - \eta_{13} n_1 e^{i2\theta_{13}} - \eta_{23} n_2 e^{i2\theta_{23}} = 0,
    \end{cases}
\end{equation}
where $\theta_{jk} = \theta_j - \theta_k$ denote the relative phase differences.  Note the phase differences must satisfy the phase closure condition, given by $\theta_{13} = \theta_{12} + \theta_{23}$. Equation~\eqref{eq:ground2} can be cast into a compact matrix form:
\begin{equation}
\mathbf{T} \mathbf{n} = -\mathbf{a},
\label{eq:ground3_1}
\end{equation}
where $\mathbf{n} = [n_1, n_2, n_3]^T$ is the superfluid density vector, $\mathbf{a} = [a_1, a_2, a_3]^T$ is the parameter vector, and $\mathbf{T}$ is the Hermitian parameter matrix:
\begin{equation}
\mathbf{T} =
\begin{bmatrix}
b_1 & -\eta_{12} e^{-i2\theta_{12}} & -\eta_{13} e^{-i2\theta_{13}} \\
-\eta_{12} e^{i2\theta_{12}} & b_2 & -\eta_{23} e^{-i2\theta_{23}} \\
-\eta_{13} e^{i2\theta_{13}} & -\eta_{23} e^{i2\theta_{23}} & b_3
\end{bmatrix}.
\label{eq:ground3}
\end{equation}

For fixed parameters and phase differences, the superfluid density vector $\mathbf{n}$ is determined by matrix inversion. To ensure the physical validity and stability of these solutions, a singularity analysis of the matrix $\mathbf{T}$ is required. The stability of the system is governed by the positive definiteness of $\mathbf{T}$ ($\mathbf{T}>0$). For this Hermitian matrix, the necessary and sufficient condition for stability is that all leading principal minors must be positive (Sylvester's Criterion\cite{Arfken2012}):
\begin{enumerate}
    \item First-order: $b_j > 0$,
    \item Second-order: $b_j b_k - \eta_{jk}^2 > 0$,
    \item Third-order: $\det \mathbf{T} >0$,
\end{enumerate}
where
\begin{equation}
\det \mathbf{T} = b_1 b_2 b_3 - b_1 \eta_{23}^2 - b_2 \eta_{13}^2 - b_3 \eta_{12}^2 - 2 \eta_{12} \eta_{23} \eta_{13}.
\label{eq:detT}
\end{equation}

In the $(\eta_{12}, \eta_{13}, \eta_{23})$ parameter space, the stability boundary defined by $\det \mathbf{T}=0$ and the second-order condition encloses a curved tetrahedron centered around the origin. This surface encloses the origin and is pinned by four discrete vertices (see Fig.~\ref{fig: tetrahedron}):
\begin{equation}
\left\{
\begin{aligned}
V_1 &= (\sqrt{b_1b_2}, \sqrt{b_1b_3}, -\sqrt{b_2b_3}) \\
V_2 &= (\sqrt{b_1b_2}, -\sqrt{b_1b_3}, \sqrt{b_2b_3}) \\
V_3 &= (-\sqrt{b_1b_2}, \sqrt{b_1b_3}, \sqrt{b_2b_3}) \\
V_4 &= (-\sqrt{b_1b_2}, -\sqrt{b_1b_3}, -\sqrt{b_2b_3}).
\end{aligned}
\right.
\label{eq:v1v2v3v4}
\end{equation}

For the symmetric case where $b_j = b$ and $\eta_{jk} = \eta$, the determinant simplifies to $\det \mathbf{T} = (b + \eta)^2(b - 2\eta)$. Combined with the second-order condition $b^2 > \eta^2$, the system remains stable within the range  $-b < \eta < b/2$.

Within the stable parameter regime, the superfluid densities are obtained via the inverse matrix:
\begin{equation}
\mathbf{n} = -\mathbf{T}^{-1} \mathbf{a}.
\label{eq:density}
\end{equation}

The explicit form of the inverse matrix is:
\begin{equation}
\mathbf{T}^{-1} = \frac{1}{\det \mathbf{T}}
\begin{bmatrix}
L_{11}  & L_{12}e^{-i2\theta_{12}} & L_{13}e^{-i2\theta_{13}} \\
L_{12}e^{i2\theta_{12}} & L_{22} & L_{23}e^{-i2\theta_{23}}  \\
L_{13}e^{i2\theta_{13}} & L_{23}e^{i2\theta_{23}} & L_{33}
\end{bmatrix},
\label{eq:invT}
\end{equation}
where
\begin{equation}
    \begin{aligned}
        L_{11} &= b_2b_3 - \eta_{23}^2, \\
        L_{22} &= b_1b_3 - \eta_{13}^2, \\
        L_{33} &= b_1b_2 - \eta_{12}^2,
    \end{aligned}
\label{eq:L_jj_definition}
\end{equation}
and
\begin{equation}
    \begin{aligned}
        L_{12} &= \eta_{12} \left( b_3 + \frac{\eta_{13} \eta_{23}}{\eta_{12}} \right),\\
        L_{13} &= \eta_{13} \left( b_2 + \frac{\eta_{12} \eta_{23}}{\eta_{13}} \right),\\
        L_{23} &= \eta_{23} \left( b_1 + \frac{\eta_{12} \eta_{13}}{\eta_{23}} \right).\\
    \end{aligned}
    \label{eq:L_jk_definition}
\end{equation}

Actually, these parameters satisfy the identity:
\begin{equation}
L_{jj} = \frac{\partial \det \mathbf{T}}{\partial b_{j}}, \quad L_{jk} = \frac{1}{2}\frac{\partial \det \mathbf{T}}{\partial \eta_{jk}}.
\end{equation}
This relation is a direct consequence of Jacobi's formula for the derivative of a determinant. Physically, $L_{jk}$ and $L_{jj}$ quantify the response of $\det \mathbf{T}$ to variations in the inter-component couplings $\eta_{jk}$ and the diagonal parameters $b_j$, respectively.

The physical requirement that the superfluid densities $n_j$ must be real-valued constrains the possible phase differences. Setting the imaginary part of the superfluid density vector to zero in Eq.~\eqref{eq:density}, $\text{Im}(\mathbf{n}) = 0$, yields the following equations:
\begin{equation}
\left\{
\begin{aligned}
    a_2 L_{12} \sin 2\theta_{12} + a_3 L_{13} \sin 2\theta_{13} &= 0 \\
    -a_1 L_{12} \sin 2\theta_{12} + a_3 L_{23} \sin 2\theta_{23} &= 0 \\
    -a_1 L_{13} \sin 2\theta_{13} - a_2 L_{23} \sin 2\theta_{23} &= 0.
\end{aligned}
\right.
\label{eq:dtheta}
\end{equation}
These equations do not explicitly contain the densities $n_j$.  This decoupling allows the phase differences $\theta_{jk}$ to be determined independently of the amplitudes, after which the corresponding densities can be calculated directly via Eq.~\eqref{eq:density}.

Equation~\eqref{eq:dtheta} always admits a set of trivial solutions satisfying $\sin 2\theta_{jk} = 0$. Since these solutions remain unchanged as system parameters vary, we refer to them as \textit{phase-locked} solutions. They correspond to 16 discrete configurations with $\theta_{jk} \in \{0, \pm\pi/2, \pi\}$ (indicated by green open circles in Fig.~\ref{solutions1}).

Provided that $L_{jk} \neq 0$ (recalling $a_j < 0$), one has from Eq.~\eqref{eq:dtheta}:
\begin{equation}
\begin{aligned}
\sin 2\theta_{12} &: \sin 2\theta_{13} : \sin 2\theta_{23} \\
&= a_3 L_{13} L_{23} : -a_2 L_{12} L_{23} : a_1 L_{12} L_{13}.
\end{aligned}
\label{eq:dtheta_ratio}
\end{equation}
This relation defines the non-trivial \textit{frustrated solutions}, characterized by phase differences that deviate from both collinear ($0, \pi$) and orthogonal ($\pm \pi/2$) alignments. These solutions form a continuous manifold where the phase differences evolve smoothly with system parameters. As long as all $L_{jk}$ remain nonzero, the above ratios can be simplified into a more transparent form:
\begin{equation}
\sin 2\theta_{12} : \sin 2\theta_{13} : \sin 2\theta_{23} = \frac{a_3}{L_{12}}  : -\frac{a_2}{L_{13}} : \frac{a_1}{L_{23}}.
\label{eq:dtheta_ratio2}
\end{equation}
highlighting the role of $L_{jk}$ as effective couplings that not only dictate the internal phase structure but also delimit the stability boundaries of the frustrated manifold.

The vanishing of any single $L_{jk}$ signals a bifurcation point where the frustrated-solution branch terminates. At these boundaries, the condition $\sin 2\theta_{jk} = 0$ is recovered, and the system reverts to the phase-locked solutions.  Consequently, the manifolds defined by $L_{jk}=0$ act as natural boundaries that partition the $(\eta_{12}, \eta_{13}, \eta_{23})$ parameter space into eight distinct sectors, as illustrated by the cross-sections in Fig.~\ref{fig:LLL}.  

These sectors consist of $R_1^{---}, R_1^{-++}, R_1^{+-+}, R_1^{++-}$, collectively referred to as the $R_1$ sectors, where the superscripts denote the signs of $(L_{12}, L_{13}, L_{23})$, together with the $R_2$--$R_5$ sectors.  Within the $R_1$ sectors, the condition, $L_{12}L_{13}L_{23} < 0$ is satisfied, whereas $L_{12}L_{13}L_{23} > 0$ holds throughout the $R_2$--$R_5$ sectors.  As demonstrated in Sec.~\ref{subsec2}, the $R_1$ sectors constitute parameter domains in which frustrated solutions can exist, whereas the $R_2$--$R_5$ sectors support only phase-locked states.

The partition induced by the $L_{jk}=0$ manifolds has important topological implications.  The frustrated solutions are structurally confined within these sectors: they form disconnected branches that cannot continuously cross the $L_{jk}=0$ boundaries.  Consequently, frustrated states remain isolated inside their respective sectors. Only at special high-symmetry points---namely the origin and the characteristic intersection points $\pm P_1$ and $\pm P_2$, where two $L_{jk}=0$ manifolds intersect---different frustrated branches can connect across neighboring sectors. The coordinates for the intersection points are given by:
\begin{equation}
\begin{aligned}
P_1 &= \left( -\sqrt{\frac{b_3}{b_2}}\eta_{12}, \quad \sqrt{b_3 b_2} \right), \\
P_2 &= \left( \sqrt{b_3 b_1}, \quad -\sqrt{\frac{b_3}{b_1}}\eta_{12} \right),
\end{aligned}
\label{eq:P1P2}
\end{equation}
which collectively define the global structure of the frustrated solutions.

Next, we determine the specific phase angles by utilizing the phase closure condition, $\theta_{13} = \theta_{12} + \theta_{23}$. To simplify the analysis, we put
\begin{equation}
\sin 2\theta_{12} = k A, \quad \sin 2\theta_{13} = k B, \quad \sin 2\theta_{23} = k C,
\label{eq:ABC}
\end{equation}
where  
\begin{equation}
A = a_3 / L_{12}, \quad B = -a_2 / L_{13}, \quad C = a_1 / L_{23},
\label{eq:ABC2}
\end{equation}
with $k$ being a scaling factor to be determined. Considering $\theta_{13} = \theta_{12} + \theta_{23}$, we obtain
\begin{equation}
k^2 = \frac{1}{C^2} \left[ 1 - \frac{(B^2 + A^2 - C^2)^2}{4A^2 B^2} \right].
\label{eq:k2}
\end{equation}
A valid frustrated solution exists only when $k^2 > 0$, which is equivalent to
\begin{equation}
    [(A+B)^2-C^2][C^2-(A-B)^2]>0.
\label{eq:k2cond}
\end{equation}
Geometrically, this condition is satisfied if and only if $A$, $B$, and $C$ satisfy the triangle inequality, i.e. 
\begin{equation}
    |A-B| < |C| < |A+B|.
\label{eq:triangleinequality}
\end{equation}
As an illustration of this geometric constraint, we provide the specific parameter regimes for $k^2 > 0$ under several typical cases in Appendix \ref{Ap:C}.

The solutions can be expressed explicitly:
\begin{equation}
    \begin{aligned}
        \cos 2\theta_{12} &= \frac{B^2 + C^2 - A^2}{2BC}, \\
        \cos 2\theta_{13} &= -\frac{A^2 + C^2 - B^2}{2AC}, \\
        \cos 2\theta_{23} &= \frac{A^2 + B^2 - C^2}{2AB}.
    \end{aligned}
    \label{eq:phase_cos}
\end{equation}
These expressions are uniquely determined by the GL parameters. By combining both the sine and cosine components, the phase angles $2\theta_{jk}$ are uniquely fixed within the $[-\pi, \pi)$ interval as follows:
\begin{equation}
    \left\{
    \begin{aligned}
        \theta_{12} &= \frac{1}{2}\operatorname{Arg} \left( \frac{B^2 + C^2 - A^2}{2BC} + ikA \right)\\
        \theta_{13} &= \frac{1}{2}\operatorname{Arg} \left( -\frac{A^2 + C^2 - B^2}{2AC} + ikB \right),
    \end{aligned}
    \right.
    \label{eq:phase_solutions_final}
\end{equation}
where the sign of $k$ determines the chirality of the solution.  This procedure yields a representative solution $P(\theta_{12}, \theta_{13})$, as indicated in Fig.~\ref{solutions1}. The other seven degenerate states are generated from $P$ through the discrete symmetry operations of the system: time-reversal symmetry ($\theta \to -\theta$, or equivalently, $k \to -k$) and the $\pi$-phase flip symmetry ($\theta \to \theta + \pi$). 

Once the relative phases $(\theta_{12}, \theta_{13})$ are obtained, the superfluid densities $n_j$ are uniquely determined in terms of Eq.~\eqref{eq:density}. Depending on the phase configurations, the solutions can be classified into five distinct categories (Cases I--V). For the frustrated solution---hereafter referred to as Case I---substituting the phase relations into Eq.~\eqref{eq:density} yields the following symmetric expressions:
\begin{equation}
\begin{aligned}
    n^\mathrm{I}_1 &= \frac{-a_1}{\det \mathbf{T}} \left( L_{11} - \frac{L_{12} L_{13}}{L_{23}} \right), \\
    n^\mathrm{I}_2 &= \frac{-a_2}{\det \mathbf{T}} \left( L_{22} - \frac{L_{12} L_{23}}{L_{13}} \right), \\
    n^\mathrm{I}_3 &= \frac{-a_3}{\det \mathbf{T}} \left( L_{33} - \frac{L_{13} L_{23}}{L_{12}} \right).
\end{aligned}
\label{eq:nj_8fold}
\end{equation}
By utilizing $L_{11} L_{23} - L_{12} L_{13} = \eta_{23} \det \mathbf{T}$ and the cyclic permutations, the expressions for the superfluid densities can be simplified to
\begin{equation}
\begin{aligned}
    n^\mathrm{I}_1 &= -\eta_{23}C = \frac{-a_1}{b_1 + \eta_{12} \eta_{13} / \eta_{23} }, \\
    n^\mathrm{I}_2 &= \eta_{13}B = \frac{-a_2}{b_2 + \eta_{12} \eta_{23} / \eta_{13}}, \\
    n^\mathrm{I}_3 &= -\eta_{12}A= \dfrac{-a_3}{b_3 + \eta_{13} \eta_{23} / \eta_{12}}.
\end{aligned}
\label{eq:nj_8fold_simple}
\end{equation}
%
%Note that a valid stationary solution of the original GL  equations ---whether corresponding to a local minimum, maximum, or saddle point--- requires $n_j^\mathrm{I} > 0$ and $k^2 > 0$, under which $n_j^\mathrm{I}$ correspond to the physical densities.  By contrast, when $k^2<0$, Eq.~\eqref{eq:nj_8fold_simple} no longer represents a physical superconducting state, but still constitutes a formal algebraic solution of the GL equations obtained after eliminating the trigonometric variables through algebraic substitutions.  For later convenience, we extend these expressions over the entire parameter space and define the corresponding \textit{virtual density} for Case I as
%
% \begin{equation}
% \boldsymbol{\mathcal{N}}^\mathrm{I} = [-\eta_{23} C, \quad  \eta_{13} B, \quad  -\eta_{12} A]^T.
% \label{eq:mathcalnj_8fold_simple}
% \end{equation}
%

For Cases II--V, the relative phases are locked at high-symmetry points, i.e., $\theta_{jk} \in \{0, \pm\pi/2, \pi\}$. The corresponding superfluid densities can therefore be directly evaluated from Eq.~\eqref{eq:density}. Below, we summarize all five possible stationary solutions of the coupled GL equations (with $l_1, l_2 \in \{0, 1\}$ and $\theta_{jk} \in [-\pi, \pi)$):
\begin{itemize}
    \item Case I: 8-fold degenerate frustrated state (TRSB). The ground-states are located at generic, non-trivial coordinates $(\theta_{12}, \theta_{13})$ given in Eq.~\eqref{eq:phase_solutions_final}, along with seven symmetry-related points generated by time-reversal symmetry (TRS) and discrete $\pi$ phase flip. The superfluid density $n^\mathrm{I}_j$ are expressed in Eq.~\eqref{eq:nj_8fold_simple}.  In this regime, the phase differences evolve continuously with the GL parameters, while remaining away from all high-symmetry points (indicated by green open dots in Fig.~\ref{solutions1}), i.e., $\theta_{jk} \not\equiv 0 \pmod{\pi/2}$.
    
    \item Case II: 4-fold degenerate phase-locked state (TRSB). The phases $(\theta_{12}, \theta_{13})$ are pinned at $(\pi/2, \pi/2) + (l_1, l_2)\pi$, and the densities are given by
    \begin{equation}
    %n^\mathrm{II}_j = \mathcal{N}_j^\mathrm{I} - \frac{\mathcal{X}^\mathrm{II}_j}{a_j} \Delta (A-B-C),
    \begin{aligned}
        n^\mathrm{II}_1 &= -\eta_{23}C + \frac{L_{12}L_{13}}{\det \mathbf{T}} (A - B - C), \\
        n^\mathrm{II}_2 &= \eta_{13}B - \frac{L_{12}L_{23}}{\det \mathbf{T}} (A - B - C), \\
        n^\mathrm{II}_3 &= -\eta_{12}A - \frac{L_{13}L_{23}}{\det \mathbf{T}} (A - B - C).
    \end{aligned}
    \label{eq:nj_II}
    \end{equation}

    \item  Case III: 4-fold degenerate phase-locked state (TRSB). $(\theta_{12}, \theta_{13})$ are pinned at $(0, \pi/2) + (l_1, l_2) \pi$, and the densities are given by
    \begin{equation}
    %n^\mathrm{III}_j = \mathcal{N}_j^\mathrm{I} - \frac{\mathcal{X}^\mathrm{III}_j}{a_j} \Delta (A+B-C),
    \begin{aligned}
        n^\mathrm{III}_1 &= -\eta_{23}C + \frac{L_{12}L_{13}}{\det \mathbf{T}} (A + B - C), \\
        n^\mathrm{III}_2 &= \eta_{13}B + \frac{L_{12}L_{23}}{\det \mathbf{T}} (A + B - C), \\
        n^\mathrm{III}_3 &= -\eta_{12}A - \frac{L_{13}L_{23}}{\det \mathbf{T}} (A + B - C).
    \end{aligned}
    \label{eq:nj_III}
    \end{equation}
    %
    % where $\mathcal{X}^\mathrm{III} = (-C, B, A)$.

    \item  Case IV: 4-fold degenerate phase-locked state (TRSB). $(\theta_{12}, \theta_{13})$ are pinned at $(\pi/2, 0) + (l_1, l_2) \pi$, and the densities are given by
    \begin{equation}
    %n^\mathrm{IV}_j = \mathcal{N}_j^\mathrm{I} - \frac{\mathcal{X}^\mathrm{IV}_j}{a_j} \Delta (A+B+C).
    \begin{aligned}
        n^\mathrm{IV}_1 &= -\eta_{23}C - \frac{L_{12}L_{13}}{\det \mathbf{T}} (A + B + C), \\
        n^\mathrm{IV}_2 &= \eta_{13}B + \frac{L_{12}L_{23}}{\det \mathbf{T}} (A + B + C), \\
        n^\mathrm{IV}_3 &= -\eta_{12}A - \frac{L_{13}L_{23}}{\det \mathbf{T}} (A + B + C).
    \end{aligned}
    \label{eq:nj_IV}
    \end{equation}
    %
    %where $\mathcal{X}^\mathrm{IV} = (C, B, A)$.

    \item  Case V: 4-fold degenerate phase-locked state (TRS). $(\theta_{12}, \theta_{13})$ are pinned at $\{ (0, 0) + (l_1, l_2) \pi$, and the densities are given by
    \begin{equation}
    %n^\mathrm{V}_j = \mathcal{N}_j^\mathrm{I} - \frac{\mathcal{X}^\mathrm{V}_j}{a_j} \Delta (A-B+C).
        \begin{aligned}
        n^\mathrm{V}_1 &= -\eta_{23}C - \frac{L_{12}L_{13}}{\det \mathbf{T}} (A - B + C), \\
        n^\mathrm{V}_2 &= \eta_{13}B - \frac{L_{12}L_{23}}{\det \mathbf{T}} (A - B + C), \\
        n^\mathrm{V}_3 &= -\eta_{12}A - \frac{L_{13}L_{23}}{\det \mathbf{T}} (A - B + C).
    \end{aligned}
    \label{eq:nj_V}
    \end{equation}
    %
    %where $\mathcal{X}^\mathrm{V} = (C, -B, A)$.
\end{itemize}

Note that $n_j^{\mathrm{II}\text{--}\mathrm{V}}$ remain analytically smooth as $L_{jk} \to 0$.  For all cases, physical stationary solutions ---whether corresponding to a local minimum, maximum, or saddle point--- require $n_j>0$, while the additional constraint $k^2>0$ applies exclusively to Case I.

%$\mathcal{N}_j^{\mathrm{II}\text{--}\mathrm{V}}$ correspond to physical stationary solutions when $n_j>0$.
%Analogously, we extend the definitions of Eqs.~\eqref{eq:nj_II}--\eqref{eq:nj_V} over the entire parameter space and introduce the corresponding virtual densities $\mathcal{N}_j^{\mathrm{II}\text{--}\mathrm{V}}$, which retain the same algebraic expressions as the original densities but are no longer constrained by the positivity condition $n_j>0$.

All five configurations can emerge as the global ground state under specific sets of GL parameters. To provide an intuitive visualization, we present typical examples of the free-energy landscapes $G(\theta_{12}, \theta_{13})$ for Cases I--V in Fig.~\ref{solutions2}. These landscapes are constructed by first determining the stationary densities $n_j(\theta_{12}, \theta_{13})$ through the equilibrium condition $\partial G / \partial n_j = 0$, and subsequently substituting them into the expression for the stationary free energy (see Appendix~\ref{Ap:B}):
\begin{equation}
    G = \frac{1}{2}(a_1 n_1 + a_2 n_2 + a_3 n_3).
    \label{eq:G_stat}
\end{equation}
As illustrated in the landscapes in Fig.~\ref{solutions2}, the global minima (ground states) are marked by white dots. The corresponding relative configurations of the three order parameters $(\psi_1, \psi_2, \psi_3)$ for each case are depicted by the vector diagrams in the upper-right panels. Notably, only Case V preserves time-reversal symmetry, whereas Cases I--IV represent distinct TRSB states.  The specific parameter regimes governing the ground state of each case will be given in Sec.~\ref{subsec2}.

% #################################################
% #########  subsection B  #######################
\subsection{Ground-state Phase Diagram}
\label{subsec2}

\begin{figure*}
    \includegraphics[width=1\linewidth]{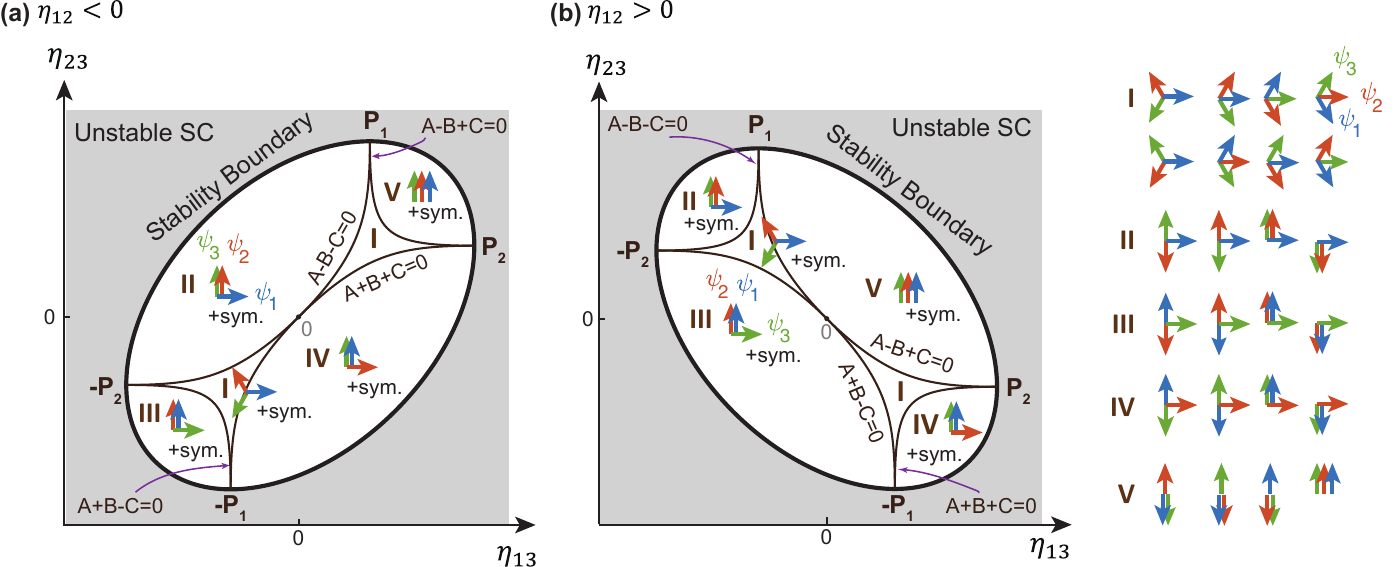}
    \caption{\textbf{Ground-state phase diagram in the $(\eta_{13}, \eta_{23})$ plane.} The diagrams are mapped for (a) a fixed repulsive coupling $\eta_{12} < 0$ and (b) a fixed attractive coupling $\eta_{12} > 0$. Regime I corresponds to the 8-fold degenerate frustrated phase, which acts as a transition zone separating four distinct 4-fold degenerate phase-locked regimes (II--V). Representative phase configurations for each regime are shown within the main plots, where "+sym." accounts for degenerate states linked by time-reversal and $\pi$-phase flip symmetries. The complete set of phase structures is fully illustrated in the right-hand insets. Superconductivity is stable within the elliptical stability boundary, defined by the positive-definiteness of the $T$-matrix ($\mathbf{T} > 0$). The analytical phase boundaries (solid brown lines) are determined by the criticality conditions $A \pm B \pm C = 0$ [Eq.~\eqref{eq:boundary_contral}]. These internal boundaries intersect the global stability limit at the points $\pm P_1$ and $\pm P_2$, with coordinates given by $P_1 = (-\eta_{12}\sqrt{b_3/b_2}, \sqrt{b_2 b_3})$ and $P_2 = (\sqrt{b_1 b_3}, -\eta_{12}\sqrt{b_3/b_1})$ [Eq.~\eqref{eq:P1P2}].
    }
    \label{fig:phaseeta}
\end{figure*}
\begin{figure*}
    \includegraphics[width=0.8\linewidth]{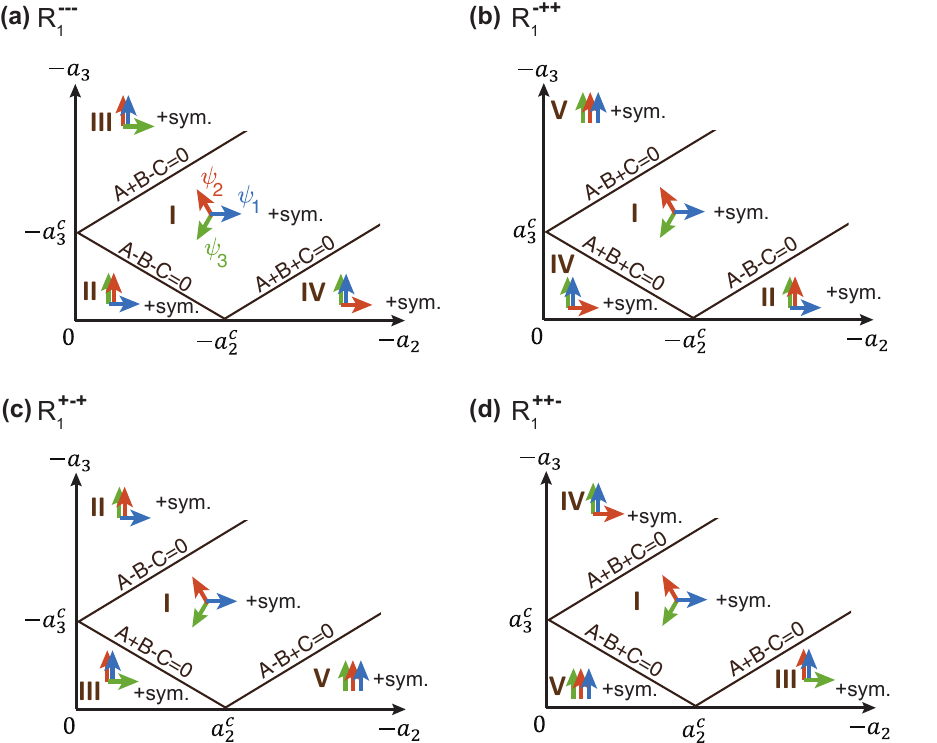}
    \caption{\textbf{Ground-state phase diagrams in the $(-a_2, -a_3)$ plane for the four $R_1$ sectors.} The phase diagrams are constructed for a fixed $a_1 < 0$ and constant values of $\eta_{jk}$ and $b_j$ within each of the four frustrated sectors: (a) $R_1^{---}$, (b) $R_1^{-++}$, (c) $R_1^{+-+}$, and (d) $R_1^{++-}$. In all four cases, Case I represents the 8-fold degenerate frustrated phase, acting as a central transition zone that separates various 4-fold degenerate phase-locked states (Cases II--V). Representative phase configurations for each state are shown in the insets, where "+sym." accounts for degenerate states linked by time-reversal and $\pi$-phase flip symmetries, following the convention in Fig.~\ref{fig:phaseeta}. The analytical phase boundaries (solid brown lines) are determined by the critical conditions $A \pm B \pm C = 0$, which appear as linear boundaries in this parameter plane and intersect the axes at critical points $|a_{2}^c|$ and $|a_{3}^c|$ [Eq.~\eqref{eq:ac}].
    }
    \label{fig:phasea}
\end{figure*}

To construct the global phase diagram, we determine the parameter regions associated with each ground-state configuration by comparing the free energies among all candidate solutions (Cases I--V).

For this purpose, we introduce the algebraically continued densities (hereafter referred to as extended densities), denoted by $\mathcal{N}_j^{\alpha}$ with $\alpha\in{\mathrm{I},\dots,\mathrm{V}}$. These quantities share the same analytical expressions as the physical densities [Eqs.~\eqref{eq:nj_8fold_simple}--\eqref{eq:nj_V}], but are formally extended over the entire parameter space by relaxing the physical constraints $n_j>0$ and $k^2>0$. The corresponding extended free-energy functional can then be written in the unified form
\begin{equation}
    \mathcal{G} = \frac{1}{2}(a_1 \mathcal{N}_1 + a_2 \mathcal{N}_2 + a_3 \mathcal{N}_3).
    \label{eq:G_stat_ext}
\end{equation}
Within the physically allowed parameter regime, $\mathcal{G}$ exactly reproduces the true free energy $G$ of the corresponding physical state.

Therefore, by evaluating and comparing $\mathcal{G}$ throughout the full parameter space, one can identify the configuration with the lowest free energy. Importantly, the resulting minimum-energy solution is found to automatically satisfy $\mathcal{N}_j>0$ within its corresponding stability region.

For Case~I, substituting the extended densities $\mathcal{N}_j^\mathrm{I}$ into Eq.~\eqref{eq:G_stat_ext} yields the extended free energy:
\begin{equation}
\mathcal{G}^\mathrm{I} = \frac{1}{2}(-a_1 \eta_{23} C + a_2 \eta_{13} B - a_3 \eta_{12} A).
\label{eq:G_I}
\end{equation}
Taking $\mathcal{G}^\mathrm{I}$ as the reference energy scale, the extended free energies of the phase-locked solutions (Cases II--V) can be expressed in compact forms:
\begin{align}
    \mathcal{G}^\mathrm{II}  &= \mathcal{G}^\mathrm{I} - \frac{1}{2} \Delta (A - B - C)^2, \label{eq:G_II} \\
    \mathcal{G}^\mathrm{III} &= \mathcal{G}^\mathrm{I} - \frac{1}{2} \Delta (A + B - C)^2, \label{eq:G_III} \\
    \mathcal{G}^\mathrm{IV}  &= \mathcal{G}^\mathrm{I} - \frac{1}{2} \Delta (A + B + C)^2, \label{eq:G_IV} \\
    \mathcal{G}^\mathrm{V}   &= \mathcal{G}^\mathrm{I} - \frac{1}{2} \Delta (A - B + C)^2, \label{eq:G_V}
\end{align}
where
\begin{equation}
\Delta = \frac{L_{12}L_{13}L_{23}}{\det \mathbf{T}}.
\label{eq:Delta}
\end{equation}
with $\Delta < 0$ in the four $R_1$ sectors, whereas $\Delta > 0$ holds throughout the $R_2$--$R_5$ sectors. Note that $\mathcal{G}^\mathrm{II}$ to $\mathcal{G}^\mathrm{V}$ remain analytically smooth in the limit $L_{jk} \to 0$.

Figure~\ref{fig:phaseeta} presents the schematic ground-state phase diagram in the Josephson-coupling parameter space.  For a fixed $\eta_{12}$, the $(\eta_{13}, \eta_{23})$ plane reveals that the 8-fold degenerate frustrated state (Case I) occupies the central region, enclosed by four phase-locked states (Cases II--V).  The analytical boundaries separating Case I from Cases II, III, IV, and V are respectively given by
\begin{equation}
    \begin{aligned}
        A - B - C &= 0, \\
        A + B - C &= 0, \\
        A + B + C &= 0, \\
        A - B + C &= 0.
    \end{aligned}
    \label{eq:boundary_contral}
\end{equation}
These boundaries are entirely contained within the four $R_1$ sectors(see Fig.~\ref{fig:NII2V}(a) for illustration). Since $\Delta<0$ in this regime, Case I is energetically favorable whenever a valid frustrated solution exists.  The existence of such a solution requires both $k^2 > 0$ and $\mathcal{N}^\mathrm{I}_j > 0$.  Since $\mathcal{N}^\mathrm{I}_j > 0$ is automatically satisfied throughout the $R_1$ sectors according to Eq.~\eqref{eq:nj_8fold_simple}, the actual ground-state domain is determined solely by the remaining condition $k^2>0$, with the phase boundaries given by Eq.~\eqref{eq:boundary_contral}.

In the frustrated Case I, the relative phases are not "pinned" to fixed values; instead, they evolve continuously as the $\eta_{jk}$ parameters vary. Upon reaching one of the analytical boundaries, the Case I solution undergoes a smooth transition into the corresponding phase-locked solution. As the system moves deeper into the phase-locked region, the inter-component coupling effectively stiffens, further stabilizing and pinning the phases into their respective locked arrangements.

Taking Case II as an exemple, the order parameters $\psi_2$ and $\psi_3$ are collinear, while both remain perpendicular to $\psi_1$. In the $2\theta$ representation, this configuration corresponds to:$$2\theta_{12} = \pi, \quad 2\theta_{13} = \pi, \quad 2\theta_{23} = 0.$$
On the boundary $A - B - C = 0$, the system reaches a critical point where the frustrated Case~I phase transitions smoothly into the phase-locked Case~II state. Moving into the Case II domain (e.g., by decreasing $\eta_{13}$) increases the repulsion between $\psi_1$ and $\psi_3$, thereby reinforcing the $2\theta_{13} = \pi$ state. Similarly, an increase in $\eta_{23}$ strengthens the $\psi_2$-$\psi_3$ attraction, while a decrease in $\eta_{12}$ enhances the $\psi_1$-$\psi_2$ repulsion. Together, these tendencies progressively lock the phases into the Case~II configuration. Consequently, Case~II occupies the upper-left side of the boundary $A-B-C=0$ in the $(\eta_{13},\eta_{23})$ plane [Fig.~\ref{fig:phaseeta}]. Moreover, $\mathcal{N}_j^\mathrm{II}>0$ is automatically satisfied throughout this region (see Appendix~\ref{Ap:C2}), confirming that Case~II indeed represents the physical ground state there.

The same mechanism applies to Cases III--V, which occupy the remaining regions surrounding the frustrated Case~I state. Energy comparisons determine the corresponding stability domains of these phase-locked states (see Appendix~\ref{Ap:C2}). Likewise, $\mathcal{N}^{\mathrm{III}-\mathrm{V}}_j > 0$ are automatically fulfilled throughout the parameter regions where Cases III--V minimize the free energy and therefore constitute the global ground states.
 
Having analyzed the effects of Josephson couplings, we now investigate the phase diagrams as a function of the parameters $a_j$. The topological behavior of these phase diagrams is fundamentally determined by the signs of $L_{jk}$, which partition the parameter space into eight distinct geometric sectors. In sectors $R_2$ through $R_5$, the system remains locked in a single ground state (Cases II through V, respectively) regardless of the values of $a_j$.

In contrast, the $R_1$ sectors exhibit a significantly richer structure. Taking the $R^{---}_1$ sector as an example, Fig.~\ref{fig:phasea}(a) displays the ground-state phase diagram in the $(-a_2, -a_3)$ plane for a fixed $a_1 < 0$, while $b_j$ and $\eta_{jk}$ are held constant. Within this regime, the system undergoes transitions between the frustrated Case I and the phase-locked Cases II, III, and IV as the $a_j$ vary.

It is important to note that while the $(\eta_{13}, \eta_{23})$ phase diagrams also contain these four states within $R^{---}_1$, the phase boundaries shift dynamically as $a_j$ are tuned. Crucially, these boundaries cannot cross into adjacent sectors because the analytical solutions are interrupted where $L_{jk} = 0$. This ensures that the phase boundaries existing in $R_1$ do not migrate into $R_2$--$R_5$, thereby preventing the emergence of alternative ground states in those regions. Consequently, the $R_1$ sector represents the physical domain where the frustrated state can emerge, a condition dictated by $L_{12}L_{13}L_{23} < 0$.

The boundaries separating Case I from Cases II, III, and IV are defined by $A-B-C=0$, $A+B-C=0$, and $A+B+C=0$, respectively. A notable distinction from the $(\eta_{13}, \eta_{23})$ plane is that the phase boundaries in the $(-a_2, -a_3)$ plane are strictly linear. These boundaries intersect the axes at the following critical values:
\begin{equation}\label{eq:ac}
\begin{aligned}
-a_2^c = -a_1\frac{L_{12}}{L_{23}}=-a_1\frac{b_3 \eta_{12} + \eta_{13} \eta_{23}}{b_1 \eta_{23} + \eta_{12} \eta_{13}}, \\ 
-a_3^c = -a_1\frac{L_{13}}{L_{23}}=-a_1 \frac{b_2 \eta_{13} + \eta_{12} \eta_{23}}{b_1 \eta_{23} + \eta_{12} \eta_{13}}.
\end{aligned}
\end{equation}

The evolution of the phase diagram reflects the underlying competition between the superconducting components, as exemplified by the $R^{---}_1$ sector in Fig.~\ref{fig:phasea}(a).  Physically, increasing $|a_2|$ enhances the $|\psi_2|$. This strengthens the effective phase repulsion between the pairs $\psi_1$-$\psi_2$ and $\psi_3$-$\psi_2$, which suppresses the frustration by driving $\psi_1$ and $\psi_3$ toward an colinear alignment ($2\theta_{13} \to 0$). Similarly, increasing the magnitudes of $|a_3|$ or $|a_1|$ strengthens their respective order parameters, driving the system toward $2\theta_{12} \to 0$ or $2\theta_{23} \to 0$.

As shown in Figs.~\ref{fig:phasea}(b)-(d), the remaining $R^{-++}_1$, $R^{+-+}_1$ and $R^{++-}_1$ sectors display similar structures. In each case, the system transitions from the frustrated Case I state to various phase-locked states as the parameters $a_j$ are tuned. Across all the frustrated $R_1$ sectors, the phase boundaries remain linear, and the physical mechanism—driven by the enhancement of specific order parameters—remains consistent.

% #################################################
% #########  subsection III  #######################
\subsection{Collective Modes and Coherence Lengths}  
\label{subsec3}

\begin{figure*}
    \includegraphics[width=0.80\textwidth]{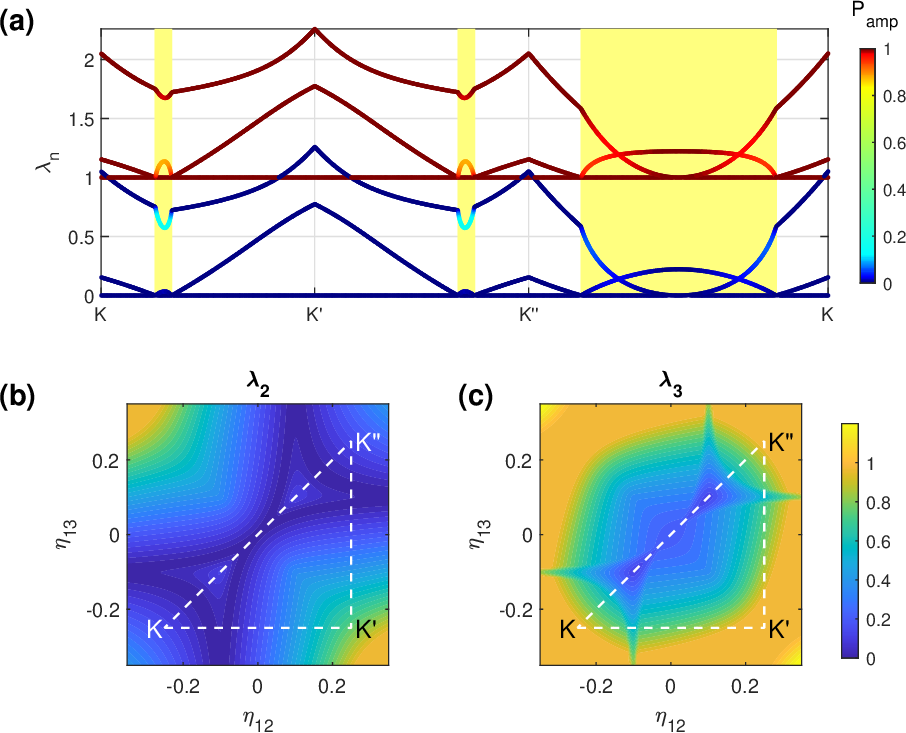}
    \caption{\textbf{Collective modes and eigenvalue spectra.}  The eigenvalues $\lambda_n$ are plotted as functions of the coupling parameters $\eta_{12}$ and $\eta_{13}$ for a fixed parameter set: $a_j = -1$, $b = 1$, $m_j = 1$, and $\eta_{23} = -0.1$. (a) Evolution of all six collective mode eigenvalues along the triangular path $K \to K' \to K'' \to K$ in the parameter plane. The vertices are defined as $K = (-0.25, -0.25)$, $K' = (0.25, -0.25)$, and $K'' = (0.25, 0.25)$. The curve colors represent the amplitude weight $P_{\text{amp}}$ of the associated eigenvectors, where red and blue indicate amplitude-dominant (Higgs-like) and phase-dominant (Leggett-like) characters, respectively. Yellow shaded regions highlight the 8-fold degenerate frustrated ground-state regimes. (b) Contour plot of the second-lowest eigenvalue $\lambda_2$ in the $(\eta_{12}, \eta_{13})$ plane. (c) Contour plot of the third-lowest eigenvalue $\lambda_3$ over the same parameter space. The dashed white triangle indicates the sampling path used in panel (a).}
    \label{fig:eigenvalues}
\end{figure*}

The stability of the identified phases can be more profoundly understood through their collective excitation spectrum; specifically, we investigate the collective modes and coherence lengths to validate the analytical phase boundaries and confirm the emergence of soft modes at the phase boundaries.

To facilitate numerical analysis, we introduce a dimensionless formulation of the free energy functional by scaling the physical quantities relative to $|a_{10}|$, $b_1$, and $m_1$ (where $a_{10} < 0$ in the superconducting state). We define the characteristic units for the order parameter amplitude $\psi_{0}$, the coherence length $\xi_0$ (defined for $\psi_1$ at $\eta_{jk}=0$), and the energy density $G_0$ as follows:
\begin{equation}
\psi_0 = \sqrt{\frac{|a_{10}|}{b_1}}, \quad \xi_0 = \frac{\hbar}{\sqrt{2m_1 |a_{10}|}}, \quad G_0 = \frac{a_{10}^2}{b_1}.
\end{equation}
The dimensionless variables are then introduced via the transformations $\mathbf{r} \to \xi_0 \tilde{\mathbf{r}}$, $\psi_j \to \psi_0 \tilde{\psi_j}$, $G \to G_0 \tilde{G}$.  Under this scaling, the gradient operator transforms as $\nabla \to \tilde{\nabla} / \xi_0$.  Substituting these into Eq.~\eqref{eq:Gall}, we obtain the dimensionless free energy density:
\begin{equation}
    \begin{split}
        \tilde{G} &= \sum_{j=1,2,3} \left[ \tilde{a_j} |\tilde{\psi_j}|^2 + \frac{\tilde{\beta}_j}{2} |\tilde{\psi_j}|^4 + \frac{1}{\tilde{m}_j} |\tilde{\nabla} \tilde{\psi}_j|^2 \right] \\
        &- \sum_{j<k} \frac{\tilde{\eta}_{jk}}{2} \left( \tilde{\psi}_j^{*2} \tilde{\psi}_k^{2} + \text{c. c.} \right), 
    \end{split}
    \label{eq:Gnodim}	
\end{equation}
where the dimensionless parameters are given by $\tilde{a}_j = \frac{a_{j0}}{|a_{10}|} (1 - T/T_c)$, $\tilde{\beta}_j = b_j/b_1$, $\tilde{m}_j = m_j/{m_1}$ and $\tilde{\eta}_{jk} = {\eta_{jk}}/{b_1}$.  In the following discussion, the tildes are omitted for brevity.

The collective modes and their associated coherence lengths are determined by analyzing the GL equations, including the kinetic energy terms:
\begin{equation}
    \begin{cases}
        a_1 \psi_1 + b_1 |\psi_1|^2 \psi_1 - \eta_{12} \psi_2^2 \psi_1^* - \eta_{13} \psi_3^2 \psi_1^* = \frac{1}{m_1} \psi_1''  \\
        a_2 \psi_2 + b_2 |\psi_2|^2 \psi_2 - \eta_{12} \psi_1^2 \psi_2^* - \eta_{23} \psi_3^2 \psi_2^* = \frac{1}{m_2} \psi_2'' \\
        a_3 \psi_3 + b_3 |\psi_3|^2 \psi_3 - \eta_{13} \psi_1^2 \psi_3^* - \eta_{23} \psi_2^2 \psi_3^* = \frac{1}{m_3} \psi_3'',
    \end{cases}
    \label{eq:glmode1}
\end{equation}
where $\psi_j''=d^2 \psi_j /d x^2$.    While we have adopted a dimensionless scaling where $b_1 = 1$ and $m_1 = 1$, these parameters are explicitly retained in Eq.~\eqref{eq:glmode1} to maintain the structural symmetry of the equations and to facilitate a clear comparison across all three components.  

To investigate the excitation spectra, we introduce small fluctuations around the ground-state order parameters:
\begin{equation}
    \psi_j = \sqrt{n_j} (1 + \delta_{2j-1}) e^{i(\theta_j + \delta_{2j})}, \quad j=1,2,3
    \label{eq:psi_delta_correction} 
\end{equation}
where $\{\delta_1, \delta_3, \delta_5\}$ and $\{\delta_2, \delta_4, \delta_6\}$ denote the amplitude and phase fluctuations, respectively.  Substituting Eq.~\eqref{eq:psi_delta_correction} into Eq.~\eqref{eq:glmode1} and linearizing with respect to the fluctuation vector $\bm{\delta} = [\delta_1, \delta_2, \delta_3, \delta_4, \delta_5, \delta_6]^\mathrm{T}$ yields the following equation:
\begin{equation}
    \mathrm{\hat{D}}\bm{\delta}=\bm{\delta''},
    \label{eq:D} 
\end{equation}
where $\mathrm{\hat{D}}$ is a $6 \times 6$ matrix (detailed in Appendix \ref{Ap:D}). Assuming a spatial dependence of the form $\delta_i = \Lambda_i e^{-x/\xi}$, Eq.~\eqref{eq:D} reduces to the eigenvalue problem:
\begin{equation}
    \mathrm{\hat{D}}\mathbf{\Lambda}=\frac{1}{\xi^2}\mathbf{\Lambda},
    \label{eq:eigen} 
\end{equation}
where the eigenvalue $1/\xi^2$ determines the characteristic coherence length $\xi$, and the eigenvector $\mathbf{\Lambda} = [\Lambda_1, \dots, \Lambda_6]^\mathrm{T}$ characterizes the nature of the collective excitation.

Fig.~\ref{fig:eigenvalues} illustrates the evolution of the eigenvalues for a representative parameter set ($a_j = -1$, $b = 1$, $m_j=1$, and $\eta_{23} = -0.1$). In Fig.~\ref{fig:eigenvalues}(a), the six eigenvalues are tracked along the $K-K'-K''$ path, with the yellow-shaded regions highlighting the 8-fold degenerate frustrated regime. The color scale indicates the amplitude weight $P_{\text{amp}}$ of each mode, definrd as:
\begin{equation}
    P_{\text{amp}}=|\delta_1|^2+|\delta_3|^2+|\delta_5|^2.
    \label{eq:eigenP} 
\end{equation}
Specifically, dark red signifies a purely amplitude-like mode ($P_{\text{amp}}=1$), whereas dark blue denotes a purely phase-like mode ($P_{\text{amp}}=0$).

For any given parameter set, the system supports three amplitude-dominated and three phase-dominated modes. Notably, a massless Goldstone mode is located at $\Lambda = 0$, reflecting the global $U(1)$ gauge symmetry. Additionally, a pure amplitude mode is pinned at $\Lambda = 1$ (due to the choice of $m_j=1$ and normalized $a_j$), which corresponds to the conventional Higgs mode.

Remarkably, the eigenvalues exhibit a distinct kink (discontinuity in slope) at the boundaries of the 8-fold degenerate frustrated phase. At these criticality points, the lowest massive (gapped) collective mode frequency undergoes significant softening, approaching zero as the system nears the phase transition. This emergence of a soft mode provides a dynamic signature of the structural instability between the frustrated and phase-locked states. Within the frustrated regime, we observe amplitude-phase mixing, a characteristic hallmark of the Higgs-Leggett mode. This hybridized collective excitation, characterized by its unique low-energy signature, is a promising candidate for experimental detection via Raman scattering spectroscopy \cite{PhysRevLett.108.177005}. The behaviors of the second- and third-lowest eigenvalues [Fig.~\ref{fig:eigenvalues}(b) and (c)] likewise display sharp changes across the phase boundaries.  Importantly, the numerically determined phase boundaries are in perfect agreement with the analytical criticality conditions derived in Eq.~\eqref{eq:boundary_contral}.

\section{Conclusion}
\label{sec:conclusion}

In summary, we have investigated a three-component Ginzburg-Landau model dominated by second-order Josephson couplings, a framework proposed to explain the $hc/(6e)$ fractional flux quantization observed in kagome superconductors\cite{Zhang_2024}. Focusing on the three-component regime, we identified five ground-state configurations: an 8-fold degenerate frustrated state (Case I) and four 4-fold non-frustrated phase-locked states (Case II--V). Due to the $\pi$-periodicity of the second-order couplings, the system inherently supports time-reversal symmetry breaking (TRSB) in all configurations except Case V.

We derived analytical expressions for these five ground states and constructed a comprehensive phase diagram by comparing their respective free energies. The stability of this frustrated state is governed by the triangle inequalities among the quantities $A$, $B$, and $C$ [Eq.~\eqref{eq:triangleinequality}].  Notably, this frustrated regime is entirely confined to the four $R_1$ sectors, where $L_{12}L_{13}L_{23} < 0$. Once the triangle-inequality conditions are violated, the system undergoes a transition into one of the phase-locked states. Consequently, the frustrated regime occupies a central region in the parameter space, separating the various non-frustrated phases.

Numerical analysis of the collective modes perfectly validates these analytical boundaries. We observed mode softening at the phase transitions boundary, providing a clear dynamical signature of the underlying instabilities. Notably, a Higgs-Leggett mode, where amplitude and phase fluctuations couple, emerges in the frustrated regime. This provides a distinct spectroscopic signature for experimental identification via Raman scattering.

As the first-order coupling is naturally suppressed in 3Q-PDW systems, this model offers a primary description for kagome superconductors such as CsV$_3$Sb$_5$. Furthermore, since parameter anisotropy is expected to emerge through strain engineering, our findings provide a guidance for the identification and manipulation of unconventional multi-component superconducting phases in quantum materials.

%%%%%%%%%  Upper: main paper  %%%%%%%%%%%%%%%%%
%%%%%%%%%  Below: appendix    %%%%%%%%%%%%%%%%%

\appendix

\renewcommand{\thefigure}{A\arabic{figure}}
\setcounter{figure}{0}

%\begin{appendix}

%
\begin{figure*}
    \includegraphics[width=0.7\textwidth]{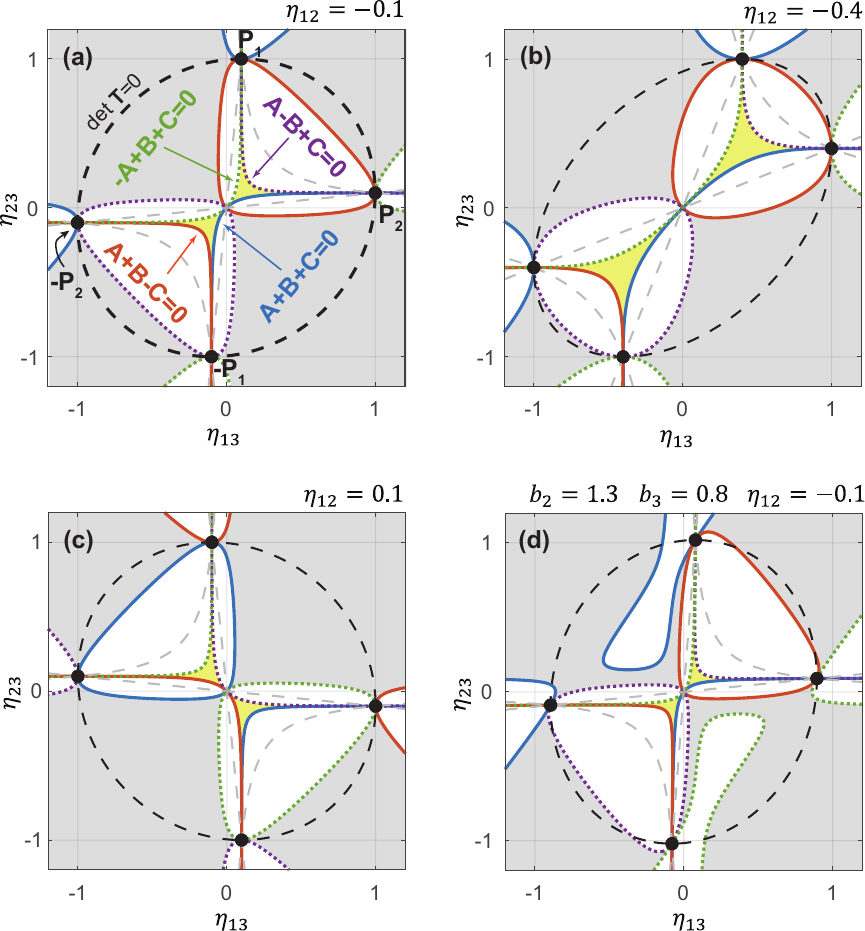}
    \caption{\textbf{Existence range of frustrated relative-phase solutions in the $(\eta_{13}, \eta_{23})$ parameter space and stability domains of frustrated solutions.} The black dashed circle delineates the stability boundary of the three-component superconducting state, as determined by the condition $\det \mathbf{T} = 0$ [Eq.~\eqref{eq:detT}]. The regions satisfying the existence condition for frustrated relative-phase solutions, $k^2 > 0$ [Eq.~\eqref{eq:k2cond}] is bounded by the curves $A \pm B \pm C = 0$ (color lines) and is represented by the combined yellow and grey areas.  Specifically, the yellow-shaded domains indicate where frustrated solutions act as stable ground states.  In contrast, grey-shaded regions indicate frustrated solutions that are physically inaccessible due to either thermodynamically unstable or negative superfluid densities ($n_j < 0$). For comparison, the $L_{jk} = 0$ discriminants from Fig.~\ref{fig:LLL} are superimposed as light-grey dashed lines. The analytical coordinates for the high-symmetry points $\pm P_1$ and $\pm P_2$ are provided in Eq.~\eqref{eq:P1P2}.  Model parameters for all panels are $a_j = -1$, with specific configurations as follows: (a) $b_j = 1, \eta_{12} = -0.1$;(b) $b_j = 1, \eta_{12} = -0.4$;(c) $b_j = 1, \eta_{12} = 0.1$;(d) $b_1 = 1, b_2 = 1.3, b_3 = 0.8, \eta_{12} = -0.1$.
    }
    \label{fig:phasek2}
\end{figure*}
\begin{figure}
    \includegraphics[width=0.8\columnwidth]{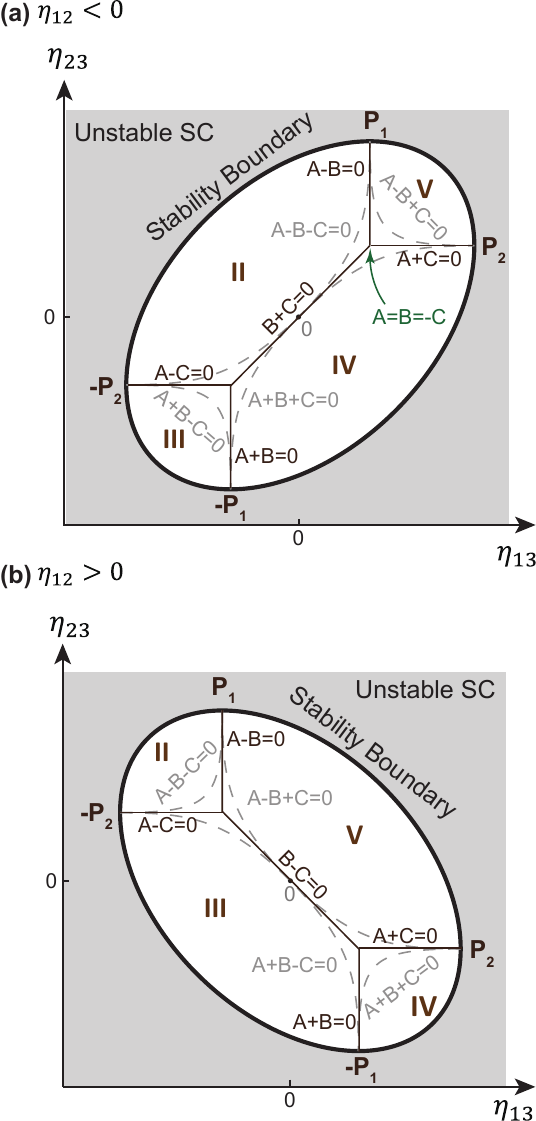}
    \caption{\textbf{Schematic representation of the Energetic competition among the four phase-locked states (Cases II--V).} Within the elliptical stability boundary defined by $\det \mathbf{T} = 0$, the regions where each phase-locked configuration possesses the lowest relative free energy are illustrated. The solid red-brown lines schematically mark the phase-transition boundaries between adjacent phase-locked domains, governed by the relations $A \pm B = 0$, $A \pm C = 0$, and $B \pm C = 0$. Although these boundaries are depicted as straight lines for conceptual clarity, the actual transition trajectories may exhibit curvature. Crucially, these interfaces are strictly confined within the sectors bounded by the grey dashed lines ($A \pm B \pm C = 0$), which delineate the physical existence and stability range of the frustrated Case~I solution. Within this central domain, the frustrated state possesses the lowest free energy and thus emerges as the true ground state. Outside this domain, the corresponding phase-locked state minimizes the global GL free-energy functional. Consequently, the phase transition between the frustrated and phase-locked regimes is strictly prescribed by the analytical boundaries $A \pm B \pm C = 0$.
    }
    \label{fig:GII2V}
\end{figure}
\begin{figure}
    \includegraphics[width=\columnwidth]{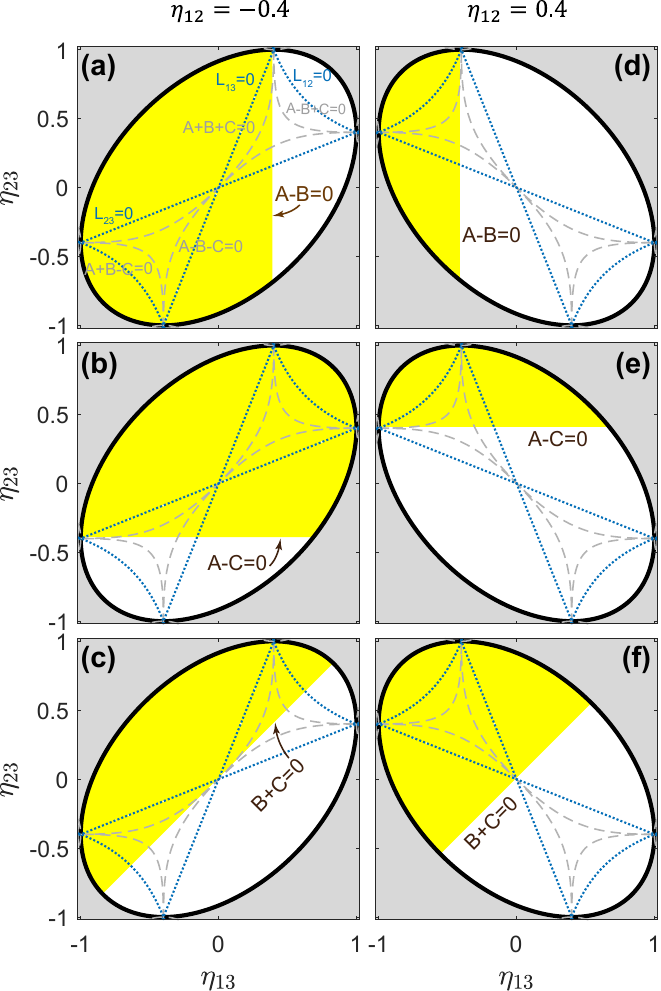}
    \caption{\textbf{Positivity of $N_j^{\mathrm{II}}$ within the region enclosed by $A-B=0$, $C-A=0$, and $C+B=0$.} The left and right columns correspond to fixed Josephson-coupling parameters $\eta{12}=-0.4$ [panels (a)-(c)] and $\eta_{12}=0.4$ [panels (d)-(f)], respectively. In each panel, the solid black ellipse denotes the stability boundary $\det\mathbf{T}=0$; the grey-shaded exterior region corresponds to $\det\mathbf{T}<0$, where the system is unstable. The yellow-shaded regions indicate the parameter domains in which the corresponding terms entering Eq.~\eqref{eq:nj_IIp} are positive:
(a), (d) $L_{12}L_{13}(A-B)>0$;
(b), (e) $L_{12}L_{23}(C-A)>0$; and
(c), (f) $L_{13}L_{23}(C+B)>0$.
Together with the strictly positive first term in Eq.~\eqref{eq:nj_IIp}, these conditions guarantee $\mathcal{N}_j^{\mathrm{II}}>0$. Consequently, all three components $\mathcal{N}_j^{\mathrm{II}}$ ($j=1,2,3$) remain positive within the region enclosed by the boundaries $A-B=0$, $C-A=0$, and $C+B=0$.  The grey dashed lines denote the boundaries $A\pm B\pm C=0$, while the blue dotted lines correspond to $L_{jk}=0$. Other parameters: $a_j=-1$ and $b_j=1$.}
    \label{fig:NII2V}
\end{figure}

\section{Parameter domain for $k^2>0$}\label{Ap:C}

Fig.~\ref{fig:phasek2} illustrates the shaded regions (comprising both grey and yellow domains) where $k^2 > 0$ [Eqs.~\eqref{eq:k2}]. These regions define the parameter space where frustrated relative-phase solutions ($\sin{2\theta_{jk}} \neq 0$) exist.  The boundaries of these shaded regions are given by $A \pm B \pm C = 0$ and are represented by colored lines. Notably, these boundaries are fundamentally quadratic equations with respect to $\eta_{jk}$. For example, when $\eta_{12}$ is held constant, each condition yields two functional branches, both of which are fully captured in Fig.~\ref{fig:phasek2}. A key topological feature is that these analytical boundaries do not cross between sectors defined by $L_{jk}=0$; instead, they are constrained to pass through high-symmetry intersection points, such as the origin and the critical points $\pm P_1$ and $\pm P_2$, with coordinates are given in Eq.~\eqref{eq:P1P2}.

While both grey and yellow regions satisfy the existence criterion $k^2 > 0$, only the yellow-shaded regions correspond to stable frustrated ground states.  As illustrated in Fig.~\ref{fig:phasek2}(a), for $\eta_{12}<0$, these stable domains are located within $R^{---}_1$ and $R^{-++}_1$. In the $R_2$ and $R_4$ sectors, the frustrated solutions are unstable (highest energy states). In contrast, within the $R_3$ and $R_5$ sectors, the solutions are physically inaccessible due to negative Cooper-pair densities, with the $n_j$ provided in Eq.~\eqref{eq:nj_8fold_simple}.

The evolution of the $k^2 > 0$ domains follows distinct patterns as system parameters vary. As $\eta_{12}$ decreases to $-0.4$ (indicating enhanced repulsion) [Fig.~\ref{fig:phasek2}(b)], the stable yellow region expands and shifts toward higher $|\eta|$ values. During this expansion, the $\det \mathbf{T} = 0$ boundary deforms into an elongated ellipse, bringing $P_1$ and $P_2$ closer together. However, as $\eta_{12}$ attains more negative values, the $\det \mathbf{T} = 0$ manifold flattens, eventually leading to a contraction of the 8-fold ground-state domain.

When $\eta_{12} > 0$, the stable frustrated solutions (yellow regions) are located within the $R^{++-}_1$ and $R^{+-+}_1$ sectors [Fig.~\ref{fig:phasek2}(c)]. Finally, although variations in $b_2$ and $b_3$ significantly deform the unstable grey regions [Fig.~\ref{fig:phasek2}(d)], the yellow ground-state regions exhibit remarkable robustness, maintaining a nearly invariant geometric form: the triconate star.

\section{The stationary free energy}\label{Ap:B} 
In this section, we derive the stationary free energy Eq.~\eqref{eq:G_stat}.  The free energy functional of Eq.~\eqref{eq:Gall} is explicitly given by
\begin{equation}
G = \sum_{j=1,2,3} \left( a_j n_j + \frac{b_j}{2} n_j^2 \right) - \sum_{\substack{j,k=1,2,3 \\ j<k}} \eta_{jk} n_j n_k \cos 2\theta_{jk}
\label{eq:Gntheta}
\end{equation}
where $\psi_j = \sqrt{n_j} e^{i\theta_j}$ and $\theta_{jk} = \theta_j - \theta_k$.

The stationary configurations of the system are determined by $\frac{\partial G}{\partial n_j} = 0$, which yields the following set of coupled equations:
\begin{align}
a_1 + b_1 n_1 - \eta_{12} n_2 \cos 2\theta_{12} - \eta_{13} n_3 \cos 2\theta_{13} &= 0 \\
a_2 + b_2 n_2 - \eta_{12} n_1 \cos 2\theta_{12} - \eta_{23} n_3 \cos 2\theta_{23} &= 0 \\
a_3 + b_3 n_3 - \eta_{13} n_1 \cos 2\theta_{13} - \eta_{23} n_2 \cos 2\theta_{23} &= 0.
\end{align}

Multiply each equation by $n_j/2$ and sum them up:
\begin{equation}
\sum_{j=1,2,3} \left( \frac{a_j}{2} n_j + \frac{b_j}{2} n_j^2 \right) - \sum_{\substack{j,k=1,2,3 \\ j<k}} \eta_{jk} n_j n_k \cos 2\theta_{jk} = 0.
\label{eq:stat_sum}
\end{equation}
Substituting Eq.~\eqref{eq:stat_sum} back into Eq.~\eqref{eq:Gntheta}, yields
\begin{equation}
G = \frac{a_1}{2} n_1 + \frac{a_2}{2} n_2 + \frac{a_3}{2} n_3.
\end{equation}

\section{Energetic Competition and Stability of the Four Phase-Locked States}\label{Ap:C2}

By evaluating the extended free-energy differences derived from Eqs.~\eqref{eq:G_II}--\eqref{eq:G_V}, the relative stability among the four phase-locked states can be determined as
\begin{align}
    \mathcal{G}^{\mathrm{II}} - \mathcal{G}^{\mathrm{III}} &= 2\Delta B(A - C), \\
    \mathcal{G}^{\mathrm{II}} - \mathcal{G}^{\mathrm{IV}}  &= 2\Delta A(B + C), \\
    \mathcal{G}^{\mathrm{II}} - \mathcal{G}^{\mathrm{V}}   &= 2\Delta C(A - B), \\
    \mathcal{G}^{\mathrm{III}} - \mathcal{G}^{\mathrm{IV}} &= 2\Delta C(A + B), \\
    \mathcal{G}^{\mathrm{III}} - \mathcal{G}^{\mathrm{V}}  &= 2\Delta A(C - B), \\
    \mathcal{G}^{\mathrm{IV}} - \mathcal{G}^{\mathrm{V}}   &= -2\Delta B(A + C),
\end{align}
where $\Delta$ is defined in Eq.~\eqref{eq:Delta}.

These relations provide a direct criterion for comparing the free-energy hierarchy of the phase-locked states, thereby determining their respective stability domains [see Fig.~\ref{fig:GII2V}]. Importantly, the boundaries between different phase-locked configurations are strictly confined within the regions enclosed by the grey dashed lines defined by $A\pm B\pm C=0$. This follows from the fact that each boundary necessarily lies between its two adjacent dashed boundaries. For example, the line $A-B=0$ is always located between the two boundaries $A-B-C=0$ and $A-B+C=0$.

As an illustrative example, consider Case~II. In this case, the positivity of the extended densities, $\mathcal{N}_j^{\mathrm{II}}>0$, can be established within the upper-left region enclosed by the three boundaries $A-B=0$, $A-C=0$, and $B+C=0$. Rewriting the extended densities Eq.~\eqref{eq:nj_II} yields
\begin{equation}
\begin{aligned}
\begin{bmatrix}
    n_1^{\mathrm{II}} \\
    n_2^{\mathrm{II}} \\
    n_3^{\mathrm{II}}
\end{bmatrix} 
&= \frac{1}{\det \mathbf{T}} \left(
\begin{bmatrix}
    -a_1 L_{11} \\
    -a_2 L_{22} \\
    -a_3 L_{33}
\end{bmatrix} 
+ 
\begin{bmatrix}
    a_2 L_{12} + a_3 L_{13} \\
    a_1 L_{12} - a_3 L_{23} \\
    a_1 L_{13} - a_2 L_{23}
\end{bmatrix}
\right) \\
&= \frac{1}{\det \mathbf{T}} \left(
\begin{bmatrix}
    -a_1 L_{11} \\
    -a_2 L_{22} \\
    -a_3 L_{33}
\end{bmatrix} 
+ 
\begin{bmatrix}
    L_{12} L_{13} (A - B) \\
    L_{12} L_{23} (C - A) \\
    L_{13} L_{23} (C + B)
\end{bmatrix}
\right).
\end{aligned}
\label{eq:nj_IIp}
\end{equation}
The first vector on the right-hand side, consisting of the diagonal elements $L_{jj}$, is strictly positive inside the stable domain. 
For the second vector, its three components are respectively positive within the sectors defined by $L_{12} L_{13} (A - B)>0$, $L_{12} L_{23} (C - A)>0$, and $L_{13} L_{23} (C + B)>0$, which visually map to the left of $A-B=0$, above $C-A=0$, and to the upper-left side of $B+C=0$ [see an example in Fig.~\ref{fig:NII2V}].Consequently, all three components satisfy $\mathcal{N}_j^{\mathrm{II}}>0$ within the upper-left region enclosed by the boundaries $A-B=0$, $A-C=0$, and $B+C=0$. Since this region fully contains the upper-left side of the boundary $A-B-C=0$, the positivity condition $\mathcal{N}_j^{\mathrm{II}}>0$ is automatically satisfied throughout the entire Case~II ground-state domain.

Using the same analysis, one can show that the extended densities $\mathcal{N}_j^{\mathrm{III}\text{--}\mathrm{V}}$ remain strictly positive within their respective domains illustrated in Fig.~\ref{fig:GII2V}. Consequently, the positivity conditions for Cases III--V are automatically satisfied throughout the corresponding ground-state regions.

Finally, it is worth noting that the boundaries $A - B = 0$, $A + C = 0$, and $B + C = 0$ intersect at the point $A = B = -C$. This intersection acts as a virtual triple point (without considering Case I) that governs the relative stability among the phase-locked states (Cases II--V). Three other virtual triple points of this nature are located at $A = C = -B$, $A = B = C$, and $A = -B = -C$ [Fig.~\ref{fig:GII2V}].

\section{Matrix Elements $\hat{D}$ for collective modes}\label{Ap:D} 

In this section, we provide the $6\times6$ dynamical matrix $\hat{D}$ governing the collective modes of a three-component superconductor. 
\allowdisplaybreaks
\begin{align*}
D_{11} &= a_1 + 3b_1 n_1 - \eta_{12} n_2 \cos(2\theta_{12}) - \eta_{13} n_3 \cos(2\theta_{13}) \\
D_{12} &= -\bigl[ \eta_{12} n_2 \sin(2\theta_{12}) + \eta_{13} n_3 \sin(2\theta_{13}) \bigr] \\
D_{13} &= -2\eta_{12} n_2 \cos(2\theta_{12}) \\
D_{14} &= 2\eta_{12} n_2 \sin(2\theta_{12}) \\
D_{15} &= -2\eta_{13} n_3 \cos(2\theta_{13}) \\
D_{16} &= 2\eta_{13} n_3 \sin(2\theta_{13}) \\
D_{21} &= -\bigl[ \eta_{12} n_2 \sin(2\theta_{12}) + \eta_{13} n_3 \sin(2\theta_{13}) \bigr] \\
D_{22} &= a_1 + b_1 n_1 + \eta_{12} n_2 \cos(2\theta_{12}) + \eta_{13} n_3 \cos(2\theta_{13}) \\
D_{23} &= -2\eta_{12} n_2 \sin(2\theta_{12}) \\
D_{24} &= -2\eta_{12} n_2 \cos(2\theta_{12}) \\
D_{25} &= -2\eta_{13} n_3 \sin(2\theta_{13}) \\
D_{26} &= -2\eta_{13} n_3 \cos(2\theta_{13}) \\
D_{31} &= -2\eta_{12} n_1 \cos(2\theta_{12}) \\
D_{32} &= 2\eta_{12} n_1 \sin(2\theta_{12}) \\
D_{33} &= a_2 + 3b_2 n_2 - \eta_{12} n_1 \cos(2\theta_{12}) - \eta_{23} n_3 \cos(2\theta_{23}) \\
D_{34} &= -\bigl[ \eta_{12} n_1 \sin(2\theta_{12}) + \eta_{23} n_3 \sin(2\theta_{23}) \bigr] \\
D_{35} &= -2\eta_{23} n_3 \cos(2\theta_{23}) \\
D_{36} &= 2\eta_{23} n_3 \sin(2\theta_{23}) \\
D_{41} &= -2\eta_{12} n_1 \sin(2\theta_{12}) \\
D_{42} &= -2\eta_{12} n_1 \cos(2\theta_{12}) \\
D_{43} &= -\bigl[ \eta_{12} n_1 \sin(2\theta_{12}) + \eta_{23} n_3 \sin(2\theta_{23}) \bigr] \\
D_{44} &= a_2 + b_2 n_2 + \eta_{12} n_1 \cos(2\theta_{12}) + \eta_{23} n_3 \cos(2\theta_{23}) \\
D_{45} &= -2\eta_{23} n_3 \sin(2\theta_{23}) \\
D_{46} &= -2\eta_{23} n_3 \cos(2\theta_{23}) \\
D_{51} &= -2\eta_{13} n_1 \cos(2\theta_{13}) \\
D_{52} &= 2\eta_{13} n_1 \sin(2\theta_{13}) \\
D_{53} &= -2\eta_{23} n_2 \cos(2\theta_{23}) \\
D_{54} &= 2\eta_{23} n_2 \sin(2\theta_{23}) \\
D_{55} &= a_3 + 3b_3 n_3 - \eta_{13} n_1 \cos(2\theta_{13}) - \eta_{23} n_2 \cos(2\theta_{23}) \\
D_{56} &= -\bigl[ \eta_{13} n_1 \sin(2\theta_{13}) + \eta_{23} n_2 \sin(2\theta_{23}) \bigr] \\
D_{61} &= -2\eta_{13} n_1 \sin(2\theta_{13}) \\
D_{62} &= -2\eta_{13} n_1 \cos(2\theta_{13}) \\
D_{63} &= -2\eta_{23} n_2 \sin(2\theta_{23}) \\
D_{64} &= -2\eta_{23} n_2 \cos(2\theta_{23}) \\
D_{65} &= -\bigl[ \eta_{13} n_1 \sin(2\theta_{13}) + \eta_{23} n_2 \sin(2\theta_{23}) \bigr] \\
D_{66} &= a_3 + b_3 n_3 + \eta_{13} n_1 \cos(2\theta_{13}) + \eta_{23} n_2 \cos(2\theta_{23})
\end{align*}

%\end{appendix}

\section*{Acknowledgments}
L.-F.Z. and X.H. acknowledge support from the Shanghai Science and Technology Innovation Action Plan (No. 24LZ1400800).

\bibliography{apssamp}% Produces the bibliography via BibTeX.
	
\end{document}